# Robust Heterogeneous Network to Support Multitasking

[1]Muhammad Shoaib Khan                                            [2] Khaled Elleithy
[1]mukhan@bridgeport.edu [2]elleithy@bridgeport.edu
Department of Electrical and Electronics Engineering
University of Bridgeport, USA.

**Abstract:** Due to emerging technology, efficient multitasking approach is highly demanded. But it is hard to accomplish in heterogeneous wireless networks, where diverse networks have dissimilar geometric features in service and traffic models. Multitasking loss examination based on Markov chain becomes inflexible in these networks owing to rigorous computations is obligatory. This paper emphases on the performance of heterogeneous wireless networks based on multitasking. A method based on multitasking of the interrelated traffic is used to attain an approximate performance in heterogeneous wireless networks with congested traffic. The accuracy of the robust heterogeneous network with multitasking is verified by using ns2 simulations.

1. INTRODUCTION

Usually we understand the word multitasking as doing two or more duties in a certain time. Multitasking involves various levels and kinds of judgment, and reasoning needed. And it depends on the tasks if they are processed simultaneously or one after another.
The essence of multitasking lies in carrying out the tasks individually and in discrete manner instead of confusing it with the simultaneous task. It is the series of events that we perform one after the other, while considering the other variable engaged in attention and information processing. However, it is necessary to spent appropriate time while switching between the tasks. Switching between tasks is a part of sequential processing of information that will be attended, processed, encoded and stored. According to [9] there is a correlation of switching between and among tasks with "attention switching". This clearly demonstrates that a feature of multitasking is change in the individual's attention and focus rather than a mere change in activities. Multiple processes and switching among doing their tasks on a processor in a constructive manner is the characterization of multitasking in computer science. To make further clear, let's suppose in the computer processor is analogous to the human brain and each program you run is a thought, switching between processes is like switching between thoughts. If you have multiple processors on your computer, you can still multitask on each one because the computer has the capacity to run myriad of processes on each processor every second.
According to [12] [14] in today's era the concurrence of activities in fostered by new technologies no matter whether it goes beyond our culture than ever before. The augmentation of such technologies is possible because of the human ability to engage in multiple tasks in a given juncture.
The latter growing significance of media for consumers to conduct their life routines and synchronize with their surrounding ambience, which has derived in escalating attention from researchers. While the novel ways of consuming media prevails, while using a medium – media multitasking – and using more than one medium at designated time. The rationale for pursuing media multitasking and multiple media use as well as results of conducting these media practices on audiences will be interrogated. The facilitating factors or impeding this conduction will be also be discovered but media multitasking; it is also expected that media multitasking and multiple media use experience will be observed with all its dimensions.
The study of **[8]** has been conducted among 8 -18 year old youth change in average amount of time spent with different media is highly differentiated while there has been a significant increase in time spent with music/audio, television content, computers and video games, whereas time spent with print media has been decreases to great extent. [11] Claiming that new communication technologies of audiovisual kinetic culture have replaced the reading culture. According to **[7]**  common media multitasker attributes; found that laptop ownership is distinguishing attribute between heavy and light media multitasker along with this the laptop does provide you with the liberty and ability to conduct plenty of other activities. In addition to this [4] [10] also believes that media ownership effects on media  multitasking and multiple media use found that ownership of certain technologies, sufficient access to the internet in the bedroom, and ability to see television while using computer.

To complete the requirements of people, heterogeneous network is one of the best solutions for multitasking. Where many cells are used, which cover the small cells area. This gives benefit to base station to do the heterogeneous characteristics in load, bandwidth, capacity and backhaul delay. At same time Mobile Station obtains traffic of diverse applications that are heterogeneous in delay and rate [3]. Each node over heterogeneous network using multitasking services performs different kinds of services simultaneously. The beauty of multitasking facilitates to users to obtain multi services efficiently.

2. RELATED WORK

Deployment of multitasking features with heterogeneous has highly enthralled different walks of the people. From other side, mobile phones owing to its moveable features and light weight have brought tremendous success in life of human beings. In modern period, every one needs to use mobile phone as magic tool to do daily routine work. We here discuss the use of heterogeneous network with characteristics.

The Authors in [15] has discussed one of the significant indicators of evaluating a performance of heterogeneous network using multitasking is the traffic modelling. A pragmatic traffic model is needed to reveal the real services and to gather an overwhelming effect in network designing steps. The traffic models can be distinguished into two categories, namely, smooth model and non-smooth model. Smooth traffic model can be further categorized into two types which are Short-Range Dependence (SRD) and Long-Range Dependence (LRD).

Quality of support in multiservice networks can be viewed as layer-by-layer issue [2] since it can be projected to different layers as different QoS related problems. The media transfer, control and management, are fundamentally discrete architecture activities as per the separation principle. In both wire and wireless IP networks, a multiservice architecture is presented for the management and control of QoS parameters.

The authors discussed the issues of multitasking over wireless network [5] .However it also has numerous technical hindrances to ever come and reduces the cost for practical applications. The authors in [6] explains the co-existence of multiple of radio access technologies RATS in the same geographical locations is one of the essential feature of next generation wireless networks (NGWN's ). A combined management of radio resources among available RAT's has been proposed for efficient radio resource utilization and enhanced QoS provisioning in heterogeneous wireless networks (HWN). Mobile terminals NGWN's have multiple interfaces and are therefore not confined to only RAT.

The paper discusses several networks to secure session between the terminal and service system over heterogeneous network and explains it is also quite hard to distinguish the normal user's access from a malicious user's attempts while using pirated personal information. An inter trust ability between terminals and the service provider system is guaranteed by secure service framework (SSF) [1]. Our contribution is completely different because we focus to determine how many sources are used for obtaining any single service over heterogeneous networks. Each node uses multiples services and maintains the memory capacity.

3. PROPOSED MODEL OF HEROGENEOUS NETWORK FOR MULTITASKING SERVICES

Our heterogeneous wireless network supports three types of media: voice, video, and data in order to develop a stochastic model for call dynamics. Though these three types of media are contemplated in the model but the model has the capacity to cater any number of network service classes.
Furthermore, in multitasking terminal call dynamics that usually corroborate voice, video, and data but can only be connected to one RAT at one point in time. The multitasking terminal is capable of supporting one, two, three or more calls at the same juncture. For instance, a user may be watching an animal documentary in shape of video while downloading a file (data) on the same multitasking terminal. Suddenly a user may receive a phone call (voice) from anyone, and all the activities are carried out seamlessly without any disruption.
Fig. 1 illustrates the some possible states of the multitasking terminal. In the initial state, the mobile terminal has no ongoing calls but the initiation of new call will make the mobile terminal to change from the initial state to another state. Likewise termination of a types of calls will cause the mobile terminal to change from one state to another.

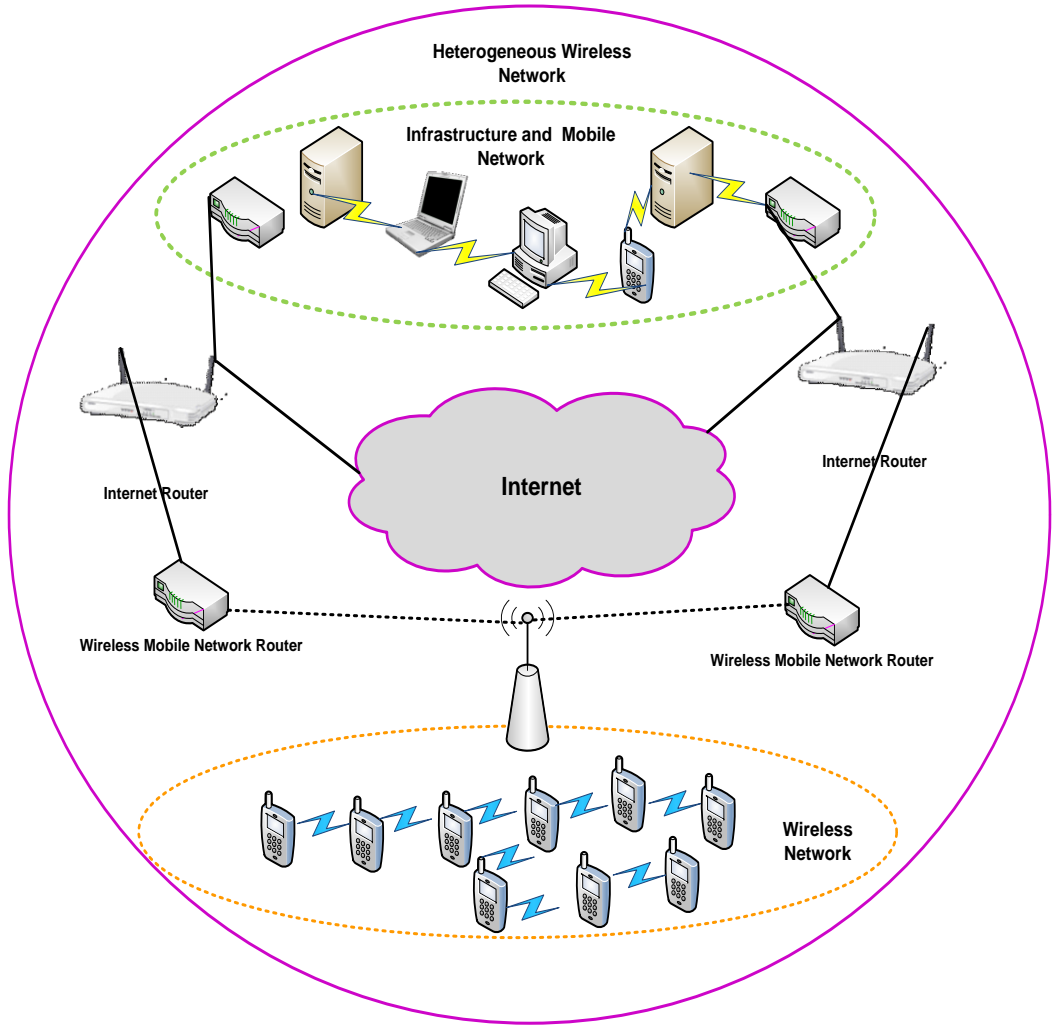

Figure 1: Heterogeneous Wireless Network Architecture

Here, we have two types of mobile nodes: static and mobility aware. Static nodes work within wired area with 8 types of activities whereas mobility nodes work with 6 types of activities, which are given in table 1.

Table 1: Multitasking functionalities performed by each mobile node in heterogeneous wireless network

| Category of node | Services used by mobile node |
|---|---|
| Wired node | Personal email, Office mail, Watch TV, Watch movies, Listen music, Play video games, Browsing and calling |
| Wireless node | Personal email, Office mail, Listen music, Play video games, Browsing and calling |

To prove these different kinds of services, we have Markov chain model that proves the validation of services performed by two different kinds of network: wired and wireless, which make heterogeneous network.

### 3.1. Markov Chain Model for Multitasking

Let $\{Y_t\}_{t=0}^{\infty}$ be an 8-Service and 6-Service Markov chain with service space S1= {0,1,2,3,4,5,6,7} and S2= {0,1,2,3,4,5}. At any distance time t, the multitasking terminal will be in state m (m Є S1} and (m Є S2}. Let D and E be an 8 * 8 and 6 *6 dimensional matrix whose (m,n)[th] element is transition probability.

$$D_{m,n} == D(Y_t = m \mid Y_{t-1} = m); \quad m,n \in S1 \quad (1)$$

$$E_{m,n} == E(Y_t = m \mid Y_{t-1} = m); \quad m,n \in S2 \quad (2)$$

At $t_{th}$ time interval, the chain builds transition from State $Y_{t-1}$ =I to state $Y_t$ in both cases ( wired nodes and wireless nodes) with probability $D_{m,n}$ and $E_{m,n}$ Let $S1^a$ ={1,2,3,4,5,6,7}  and $S2^b$ ={1,2,3,4 }   denote the set of active services of the multitasking done by wired and wireless mobile nodes in heterogeneous network.

Let us to prove using probability matrix transition, S1 and S2 are given as follows

$$D = \begin{pmatrix} D\infty & \cdots & D17 \\ \vdots & & \vdots \\ D70 & \cdots & D77 \end{pmatrix} \quad (3)$$

$\sum_{n=0}^{7} D\,m,n = 1, \quad \forall\, m \in S1,\ 0 \leq D\,m,n \leq 1$

The steady state probability of multitasking terminal S1 is obtained using following equation.

$$\pi = \pi D \quad (4)$$

Where $\pi$ is state probability vector obtained by:

$$\pi = [D_0, D_1, D_2, D_3, D_4, D_5, D_6, aD_7]$$

Thus

$$\sum_{y-0}^{7} D_y = 1 \quad (5)$$

$$E = \begin{pmatrix} E\infty & \cdots & E14 \\ \vdots & & \vdots \\ D40 & \cdots & D44 \end{pmatrix} \quad (6)$$

$\sum_{n=0}^{7} E\,m,n = 1, \quad \forall\, m \in S2,\ 0 \leq E\,m,n \leq 1$

The steady state probability of multitasking terminal S2 is obtained using following equation.

$$\pi = \pi E \quad (7)$$

Where $\pi$ is state probability vector obtained by:

$\pi = [E_0, E_1, E_2, E_3, E_4]$

$$\sum_{y-0}^{7} E_y = 1 \quad (8)$$

### 3.2. Working process of Multitasking

The flowchart / algorithm begins by checking if the task needs multitasking, it will move to the next step to request the implementation of a specific services during multitasking; and if does not require multitasking the command will be transferred back to the beginning of the algorithm/flowchart. Assuming that the task requires multitasking, it will be processed by requesting implementation of a specific services and then the command will move to the next step to check if the resource requirement for multitasking task exceeds the resource limitation, if 'Yes' the command will search for the lowest priority task. And if requirement does not exceed the given resources, the command will go to another step which will implement specific services for the given task in multitasking mode and then end the algorithm.

Continuing with search for the lowest priority task, after finding the lowest priority task the command moves onto a step where it is decided if there are more than one lowest priority task running. If the result at this step is true, the command jumps to next step to determine the task using most of the memory. And if there is only one task with the lowest priority, the command is transferred to a step, which asks the user if the lowest priority task needs to be terminated. Continuing with more than one lowest priority task and if the result is a 'Yes', then the next step will be to find the task which will be using more resources. In case, the result is 'No', even then the command will move to the next step to determine the task with higher resource requirement.

After doing so, and moving to the next step, the command will ask for termination of task. If it's a 'Yes', the task will be terminated in the next step and command will be transferred back to an earlier step which will find out if the task is beyond resources. If it's 'No', the specific multitasking process will finish given in figure 2.

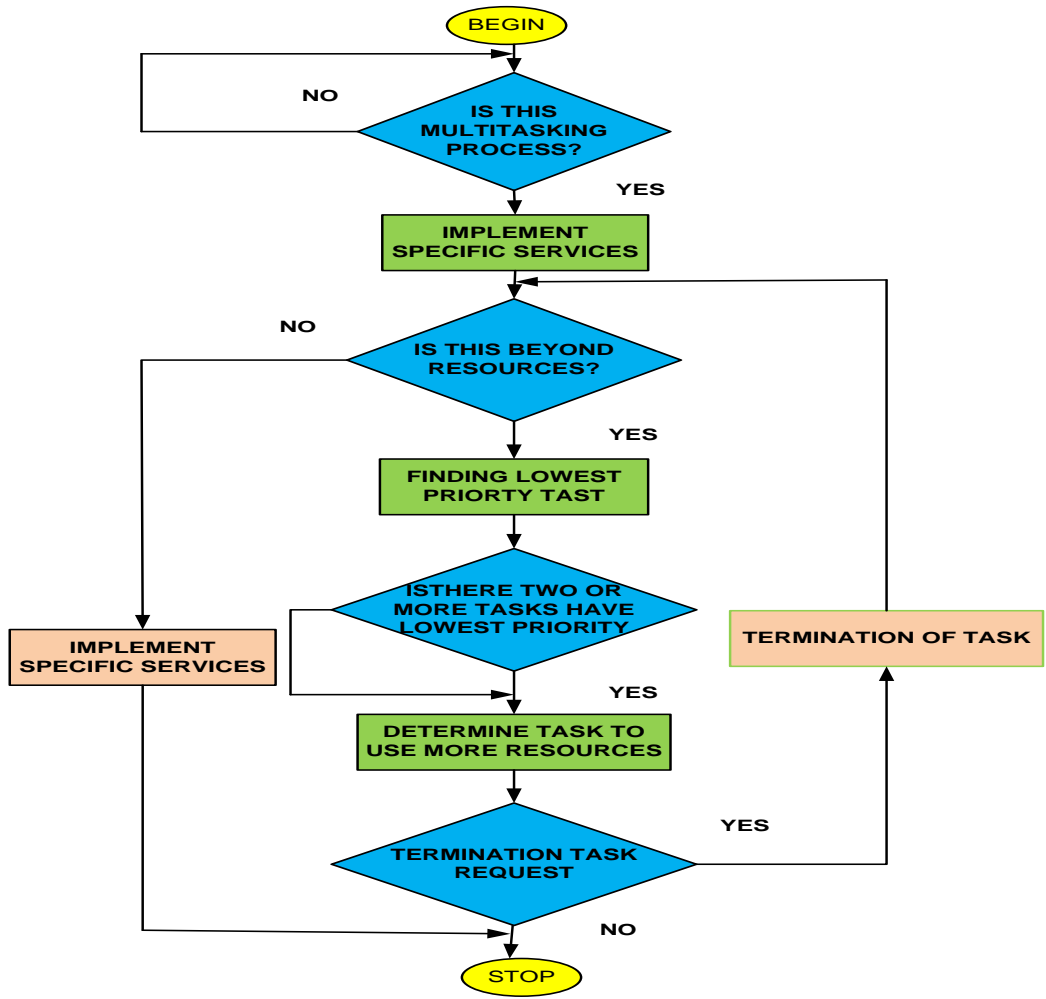

Figure 2: Working process of multitasking

3.3. Realistic Scenario for multitasking

We deploy the heterogeneous network with multitasking features in realistic scenario that is mapped in a university. This network may be connected to different peripherals/terminals/nodes that may be wired or wireless and requires different set of resources to perform certain functions. The network of information technology (IT) department is connected to the internet and the network is connected to the on-campus labs, offices, library, security office and recreational activities centers, dorm etc. These connections can be wireless or wired. All these facilities are being provided under single network and hence there is a very strong desire for multitasking. This heterogeneous network connects the infrastructure and mobile in a wired manner because we do not need to establish infrastructure everywhere in university.

It saves resources and provides cheaper services to all users. All of the peripherals/terminals/nodes are also connected to the internet routers via wire. Similarly the internet router are connected to the internet in a wired fashion. This wireless access point network for wireless connections can be utilized in many ways. The security office can communicate by using walkie talkies, the personal alarm locator (PAL) are significant part of security provider for the staff and on-campus students in university. The Wi-Fi hotspots are connected to one central server for providing data, voice and video services. The labs require so many different resources for performing tasks. The security cameras (CCTV) are connected to the same network for surveillance of different parts of on-campus and storage of data at servers.

Desk phones can also be wireless and connected to the network and advantage of being wireless would be, easily relocate, reduction of network cables, cost efficient and above all; all this is happening under one storage server. Distance learning programs can also be offered under the same server. Servers will be able store and subsequently broadcast the lecture accordingly. As it is very clear now that heterogeneous network provides many multitasking support to several locations by saving many and resources.

4. SIMULATION SETUP AND ANYLYSIS OF RESULTS

We have developed simulation model by using network simulator version 2.34 (NS2) [13] with IEEE 802.11b and data rate is 54 Mb/Sec underlying routing protocol is temporally ordered routing algorithm (TORA). The advantage of using TORA is easy to control with on-demand routing features. TORA also limits the propagation of control messages. The routes are maintained efficiently. The topology consists of wired and wireless network. The wired links use high bandwidth with trivial delay such that end-to-end delay is mostly dependent on performance of selected wireless access network. After warm up time of 30 seconds the wireless node attempts to send the packets with multitasking services. The packets are then sent using FTP application. The used parameters for this simulation are shown in table 2.

Table 2: Showing used parameters in simulation

| | |
|---|---|
| Debug Mask | 1 |
| Debug File Index | 0 |
| MTU | 1500 |
| Data Chunk Size | 512 and 1468 |
| Number of out streams | 1 |
| CMT congestion window | 1 |
| CMT Del Acknowledgement | 1 |
| RTX congestion window | 4 |
| Heart Beat Timer | 0 |
| Initial Receiving window | 65536 |
| Queue size limit | 50 |
| Maximum Initial Retransmits | 9 attempts |
| RTO Min | 1second |
| RTO Max | 60 seconds |
| RTO Initial | 4 seconds |
| Router | TORA |
| RTO Beta | 1/4 |
| RTO Alpha | 1/8 |
| Path Maximum Retransmission | 6 attempts (per destination address) |
| RTO Initial | 4 seconds |
| Application Buffer size | 0 |
| Send Buffer Size | 0 |
| Channel type | Wireless Channel |
| Drop Tail | 5 Mb 200ms |
| Simulation time | 400 seconds |
| Packet size | 1024 bytes |
| Application | ftp |
| Burst time | 0.5 second |
| Data rate | 54 Mb/Sec |
| No: of changes | 10 |
| Radio-propagation model | Two Ray Ground |
| Network interface type | OFDM |
| MAC type | Mac/802_16/BS |
| Link Layer type | Logical Link |
| Interface queue type | Drop Tail/Priority Queue |
| Pause time | 3 seconds |
| Transport level protocol | Standard TCP |

We hereby compare the performance of multitasking and without multitasking features. Figure 3 shows the Average throughput of mobile nodes during the multitasking and without multitasking services. The throughput of multitasking is higher than without multitasking scheme because nodes have capability to do different tasks simultaneously that is reason, high throughput is achieved using multitasking. Figure 3.

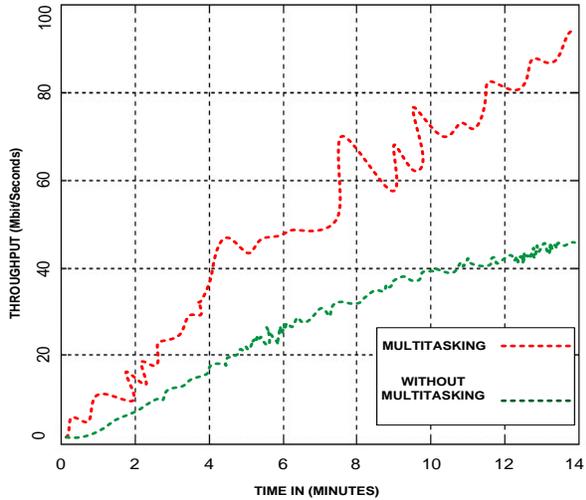

Figure 3: Throughput of multitasking and without multitasking and without multitasking

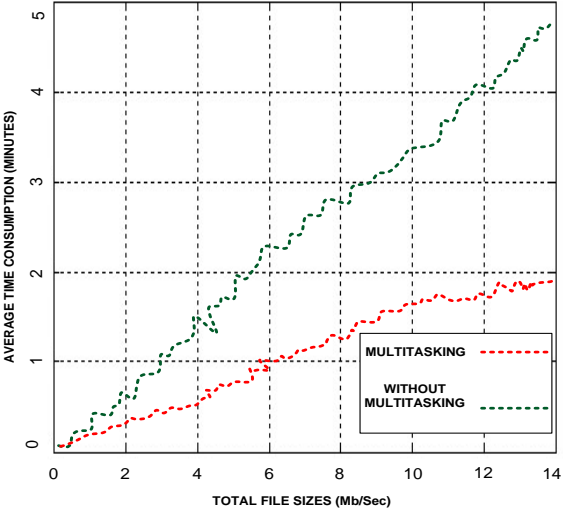

Figure 4: Consumption of time for transfer of files with multitasking at different time interval

Figure 4 shows average completion time for transfer of all sent files. For nodes, using multitasking services take lower completion time. As estimated average completion time for transfer of the files increases with increase of the file size. Initial results demonstrate that the proposed robust multitasking strategy yields considerably better results for all the nodes over all transferred of all file sizes.

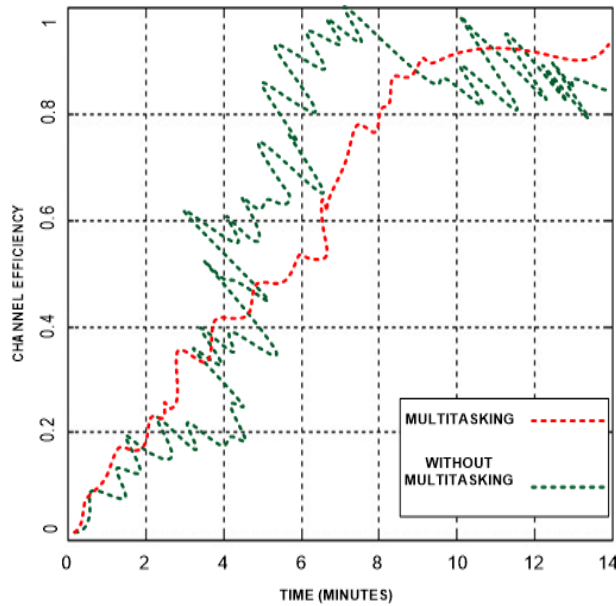
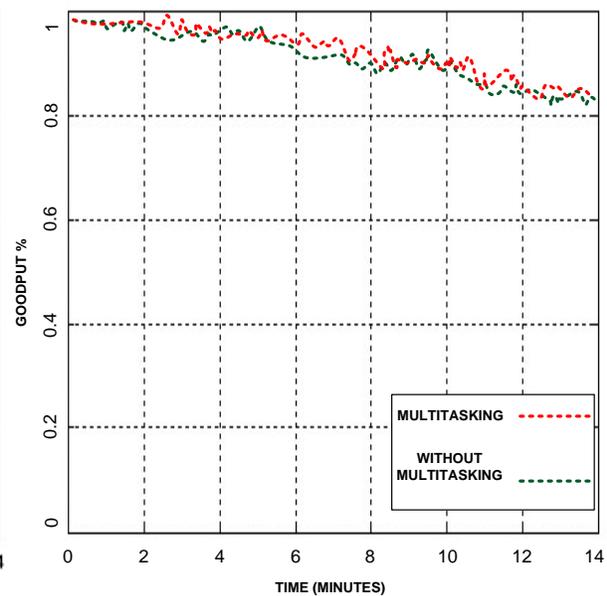

Figure 5: Channel utilization with multitasking and without multitasking

Figure 6: Goodput % with multitasking and without multitasking at different time interval

Throughput is sometimes standardized and measured in form of percentage, but standardization may cause misperception concerning what the percentage is associated to. Channel efficiency, Channel utilization and packet drop rate are less vague terms in percentage. The performance of channel efficiency is considered as bandwidth utilization efficiency. The achieved throughput of channel efficiency in percentage is related to bitrate of network in form of bit/sec of communication channel. For instance, throughput is 80 Mbit/sec in 100 Mbit/sec Ethernet connection. The channel efficiency is counted 80%. The channel efficiency calculates not only data bits but also overhead of the channel. Here, the figure 5 shows the channel efficiency of nodes using multitasking and without multitasking techniques. The channel efficiency with multitasking is almost uniform whereas the channel efficiency without multitasking is variable. It is noted that channel efficiency cannot highly be affected due to multitasking technique even nodes do many tasks simultaneously.

We here discuss the goodput of multitasking and without multitasking in figure 5. The goodput is considered as throughput of application level. It can be measured with delivered amount of data in form of bits by network to certain destination in unit per time. The goodput of multitasking and without multitasking is relatedly similar even more throughput is obtained with multitasking by performing several tasks simultaneously. We here excludes protocol overhead and retransmitted data packets in goodput. The goodput is based counted with amount of time from first packet is sent until last packet is delivered given in the following formula.

Goodput= Number of acknowledged packets / Number of transmitted packets * 100.

From the results, it is validated that performance of multitasking is not affected by getting maximum amount of throughput. Multitasking provides the better option to perform several tasks for saving the time and resources.

CONCLUSION

The robust heterogeneous network with multitasking support has been introduced for delivery of maximum throughput. The proposed model for heterogeneous network will be benefited for connecting different areas where is not infrastructure based technology is existing. In addition several tasks can be performed by several locations simultaneously to save time and resources. The proposed work has been supported with Markov chain model for multitasking that gives accurate time for performing all activates either wired or wireless section of heterogeneous of network.  On basis of simulation done in ns2, we validate and prove the strength of our proposed multitasking approach over heterogeneous network because we have obtained better throughput as compare without multitasking approach even consume similar amount of bandwidth and channel utilization. Multitasking approach can be deployed in several applications for instance business and communication while performing many tasks without consumption of maximum resources in short period of time. In future, we will simulate highly congested network by increasing the number of mobile nodes to examine the performance of multitasking over heterogeneous network.

**BIOGRAPHY:**

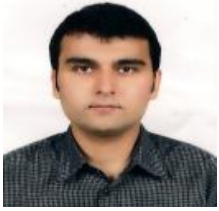

**Mr. Muhammad Shoaib Khan** received his BS degree in Computer Engineering from Sir Syed University of Engineering & Technology, Karachi, Pakistan in 2008. He has currently been involved in his Master dissertation from Electrical and Electronics Engineering department at University of Bridgeport, Connecticut, USA. His research interests include Wireless design network and routing protocols. **Mr. Muhammad Shoaib Khan** may be reached at mukhan@bridgeport.edu.

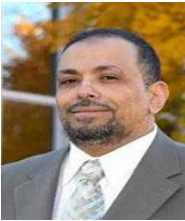

**Dr.** Elleithy is the Associate Dean for Graduate Studies in the School of Engineering at the University of Bridgeport. He has research interests are in the areas of network security, mobile communications, and formal approaches for design and verification. He has published more than two hundred research papers in international journals and conferences in his areas of expertise.

Dr. Elleithy is the co-chair of the International Joint Conferences on Computer, Information, and Systems Sciences, and Engineering (CISSE). CISSE is the first Engineering/Computing and Systems Research E-Conference in the world to be completely conducted online in real-time via the internet and was successfully running for four years.

Dr. Elleithy is the editor or co-editor of 10 books published by Springer for advances on Innovations and Advanced Techniques in Systems, Computing Sciences and Software. Dr. Elleithy has more than 20 years of teaching experience. He supervised hundreds of senior projects as well as MS theses. He developed and introduced many new undergraduate/graduate courses. He also developed new teaching / research laboratories in his area of expertise. Dr. Elleithy is the recipient of the 2006 - 2007 University of Bridgeport Professor of the Year Award.

Dr. Elleithy received the B.Sc. degree in computer science and automatic control from Alexandria University in 1983, the MS Degree in computer networks from the same university in 1986, and the MS and Ph.D. degrees in computer science from The Center for Advanced Computer Studies in the University of Louisiana at Lafayette in 1988 and 1990, respectively. Dr Elleithy may be reached at elleithy@bridgeport.edu.